\documentstyle[prl,twocolumn,aps]{revtex}
\input{epsf}
\begin{document}
\draft
\twocolumn[\hsize\textwidth\columnwidth\hsize\csname @twocolumnfalse\endcsname
\title{3D Anderson transition for two electrons in 2D}

\author{D. L. Shepelyansky}

\address {Laboratoire de Physique Quantique, UMR 5626 du CNRS, 
Universit\'e Paul Sabatier, F-31062 Toulouse Cedex 4, France}

\date{February 17, 1999}

\maketitle

\begin{abstract}
It is shown that the Coulomb interaction can lead to
delocalization of two electron states in two-dimensional (2D)
disordered potential in a way similar to the Anderson
transition in three dimensions (3D). At fixed disorder strength
the localized phase corresponds to low electron density 
and large value of parameter $r_s$.

\end{abstract}
\vskip1pc]

\narrowtext


Contrary to the well established theoretical result \cite{and79},
according to which noninteracting electrons are always
localized in 2D disordered potential, the pioneering
experiment by Kravchenko {\it et al.} \cite{krav94}
demonstrated the existence of metal-insulator transition
for real interacting electrons in 2D. The ensemble of
experimental data obtained by different groups 
\cite{krav96,popovic,canada,yael,alex,pudalov}
clearly indicates the important role played by interaction.
In the majority of experiments the Coulomb energy
of electron - electron interaction $E_{ee}$
is significantly larger than the Fermi energy
$E_F$, estimated for noninteracting electron gas 
in absence of disorder. The ratio of these energies is characterized
by the dimensionless parameter $r_s = 1/{\sqrt{\pi n_s} {a^*_B}} \simeq
E_{ee}/E_F$, where $n_s$ is the electron density in 2D,
and $a^*_B = \hbar^2 \epsilon_0/m^* e^2$, $m^*$, $\epsilon_0$
are the effective Bohr radius, electron mass and dielectric constant
respectively. Such large $r_s$ values as 10 - 30
have been reached experimentally 
\cite{krav94,krav96,popovic,canada,yael,alex,pudalov}.
At these $r_s$ the electrons are located far from each 
other and it is natural to assume that in this regime the
interaction effects will be dominated by pair interaction.
The important role of the residual two-body interaction is also
clear from the fact that in the Hartree-Fock 
(mean field) approximation the problem is
again reduced to the one-particle 2D disordered potential 
with localized eigenstates \cite{and79}.

The problem of two electrons interacting in the localized
phase is rather nontrivial. Indeed, recently it has been
shown that a short range repulsive/attractive interaction
between two particles can destroy one-particle
localization and lead to creation of 
pairs propagating on a distance much larger than their
size \cite{ds94,imry,pichard,oppen,moriond}. 
The pair size is of the order of one-particle localization
length $l_1$. Inside this length the collisions between
particles destroy the quantum interference 
that results in their coherent propagation on 
a distance $l_c \gg l_1$. The important point
is that only pairs can 
propagate on a large distance. Indeed, the 
particles separated by a distance $R \gg l_1$
have exponentially small overlap, the
interaction between them is weak and
such states are localized as in the noninteracting case.
According to the theoretical
estimates \cite{ds94,imry,moriond} in 2D the localization length 
$l_c$ grows exponentially with $l_1$ according to the
relation $\ln (l_c/l_1) \sim \kappa > 1$.
Here $\kappa \sim \Gamma_2 \rho_2$, where $\Gamma_2 \sim U^2/(V l_1^2)$
is the interaction induced transition rate between localized states
in {\it e.g.} 2D Anderson model,
$\rho_2 \sim l_1^4/V$ is the density of two-particle states
directly coupled by interaction,
$V$ is the hopping between nearest sites, $U$
is on (nearest) site interaction, and energy is taken in the 
middle of the band. In a sense the above estimate is
similar to the case of one-particle localization in 2D
where $\ln l_1 \sim k_F \ell \sim (V/W)^2 $
and the product of the Fermi wave vector $k_F$
on mean free path $\ell$ is proportional to a local diffusion rate
\cite{rmph}; $W$ is the strength of on site disorder.
Indeed, in the same manner the interaction induced diffusion rate of a
pair is given by $D_2 \sim l_1^2 \Gamma_2 \sim \kappa/l_1^2 \propto \ln l_c$.
According to the above estimates $l_c$ should vary smoothly
with the effective interaction strength characterized by 
the dimensionless parameter $\kappa$. However, this
consideration is valid only for a short range interaction
while the analysis of the long range Coulomb interaction
requires a separate study. The investigation of
this case is also dictated by the experiments 
\cite{krav94,krav96,popovic,canada,yael,alex,pudalov}
where the electrons are not screened and are located
far from each other $(r_s \gg 1)$. On a qualitative grounds
one can expect that the effect of Coulomb interaction
will be stronger since electrons are always interacting
in a difference from the case of short range interaction.
As we will see later the interaction effects will play an
important role even at low density
when the electrons are far from each other $(R \gg l_1)$
and where the interaction can lead to the delocalization
transition similar to one in the 3D Anderson model.
It is convenient to study this transition by the means
of level spacing statistics as it was done for 3D 
one-particle case in \cite{shklov}.

To analyze the effect of Coulomb interaction between
two electrons let us consider the 2D Anderson
model with the diagonal disorder $(-W/2 < E_i < W/2)$,
hopping $V$, the lattice constant 
$a=1$ and the interaction $U/|{\bf r}_1 - {\bf r}_2|$.
In these notations 
$r_s=U/(2V\sqrt{\pi n_s})$ and 
it is convenient to introduce another
dimensionless parameter $r_L = U l_1/2\sqrt{\pi}V$
which is equal to $r_s$ value at $n_s=1/l_1^2$.
We will 
consider the case with $U \sim V$ and $r_s \gg 1$  when
the average distance between electrons $R=|{\bf r}_1-{\bf r}_2|$
is much larger than their noninteracting localization length:
$R \sim 1/\sqrt{n_s} \sim r_s \gg l_1 \gg 1$. 
In this case the two-body interelectron interaction
has a dipole-dipole form and is of the order of
$U_{dd} \sim U {\Delta {\bf r}_1} {\Delta {\bf r}_2}/R^3 \sim
U l_1^2/R^3$. Indeed, the first two terms in the expansion of Coulomb
interaction give only  mean field corrections to one-particle
potential and the nontrivial two-body term appears only
in the second order in the electron displacements
${\Delta {\bf r}_1} \sim {\Delta {\bf r}_2} \sim l_1$
near their initial positions 
${{\bf r}_{1,2}}$.
The matrix element of this dipole-dipole interaction
between localized noninteracting eigenstates
is of the order of
$U_s \sim U \sum {{\Delta {\bf r}_1} {\Delta {\bf r}_2}} \psi^4/R^3
\sim U/R^3$. Here 
$\psi \sim \exp(-| {\Delta {\bf r}_{1,2}}|/l_1)/ l_1$
are localized one-electron states and due to localization
the sum runs over $l_1^4$ sites and each term in the sum has a random sign.
According to the Fermi golden rule these matrix
elements give the interaction induced transition rate
$\Gamma_{e} \sim U_s^2 \rho_2 \sim U_{dd}^2/V$,
where the density of coupled states in the middle of the
energy band is still $\rho_2 \sim l_1^4/V$ since 
due to localization only jumps on a distance $l_1$ are allowed.
These interaction induced matrix elements
mix two-electron states if $\kappa_e \sim \Gamma_e \rho_2 > 1$,
that corresponds to  $R < l_1 (U l_1/V)^{1/3}$
(a similar estimate for electrons in 3D was given in Ref. 1b).
Since $l_1 \gg 1$ the condition $R \gg l_1$ is
still satisfied. 
For $\kappa_e > 1$ these transitions lead to a diffusion with the rate 
\begin{equation}
\label{dif}
D_e \sim l_1^2 \Gamma_e \sim V \kappa_e/l_1^2 
\end{equation}
This diffusion expands  in an effective 3D space. Indeed, the center of mass
of two electrons diffuses in 2D lattice plane
and in addition the electrons diffusively rotate
on a ring of radius $R$ and width $l_1$. The radius of the
ring is related to the {\it e-e}-energy
$E \sim U/R$ which remains constant. Since $R \gg l_1$
it takes a long time to make one rotation along the ring.
As for the 3D Anderson model this diffusion
becomes delocalized when the hopping is larger than
the level spacing between directly coupled states,
namely:
\begin{equation}
\label{deloc}
\chi_e \sim \kappa_e^{1/6} \sim r_L^{4/3}/r_s > 1
\end{equation}
Formally the situation corresponds to a quasi-two dimensional
case with $M_{ef} \approx \pi R/l_1 = \pi r_L^{1/3} >> 1$
parallel planes (number of circles of size 
$l_1$ in the ring) so that the pair localization length $l_c$
jumps from $l_c \sim l_1$ for $\kappa_e <1$ to
$l_c \sim l_1 \exp(\pi \kappa_e r_L^{1/3}) \gg l_1$
above the transition $\kappa_e > 1$. The transition is sharp and
similar to 3D Anderson transition when $r_s > r_L \gg 1$.
If electrons would be able to move inside the ring then $M_{ef}$
would be even larger $(M_{ef} \sim r_L^{2/3})$.

It is important to stress that the parameter $\chi_e$
which determines the delocalization border
and measures the effective strength of two-body
interaction decreases with the increase of $r_s$.
Apparently, this looks to be against the common lore
according to which the larger is $r_s$ the stronger
is the {\it e-e}-interaction. The reason of this contradiction
with (\ref{deloc}) is simply due to the fact that
$r_s$ compares $E_{ee}$ with $E_F$ computed in the {\it absence}
of disorder. In the presence of not very weak disorder
($r_D = E_{ee}/W \ll 1$ and $r_L \gg 1$) the one-electron states
are localized and form the basis of Coulomb glass \cite{efros}.
In this Coulomb glass phase the  {\it e-e}-interaction
becomes weaker and weaker with the growth of average
distance between electrons $R \sim n_s^{-1/2} \propto r_s$
in the natural agreement with (\ref{deloc}). The transition
border (\ref{deloc}) was obtained for excited states.
However, it is clear that if the interaction is not able
to delocalize the excited states then the low energy states 
will also remain localized since two-electron density $\rho_2$
drops at low energy. In this sense (\ref{deloc})
determines the upper border for $r_s$.

To study the delocalization transition (\ref{deloc})
the level spacing statistics $P(s)$ is determined numerically
for different system sizes $L$. To follow the transition
from the localized phase with the Poisson statistics $P_P(s)$
to delocalized one with the Wigner-Dyson statistics $P_{WD}(s)$
it is convenient to use the parameter 
$\eta=\int_0^{s_0}
(P(s)-P_{WD}(s)) ds / \int_0^{s_0} (P_{P}(s)-P_{WD}(s)) ds$,
where  $s_0=0.4729...$ is the intersection point of $P_P(s)$ and $P_{WD}(s)$
\cite{jac2}. In this way $\eta=1$
corresponds to the Poissonian case, and $\eta$=0 to $P_{WD}(s)$.
The dependence of $\eta$ on the one electron 
energy $\epsilon = E/2$, counted from the ground state
is shown in Fig. 1 for different disorder $W$ and interaction strength $U$.
Usually ND=4000 realizations of disorder are used and 
in addition the average in a small energy interval
allows to increase the total statistics for $P(s)$ and 
$\eta$ from $NS=12000$
for low energy states up to $NS=10^6$ at high energies
with larger density of levels. The matrix diagonalization
is done in the one-electron eigenbasis truncated at high 
energies that allowed to study two electron low energy 
excitations (with energy $E$) at large system sizes $L \leq 24$.
The periodic boundary conditions are used for one-electron states,
the Coulomb interaction is taken between electrons in
one cell of size $L$ and with 8 charge images
in nearby 8 cells.
The Coulomb interaction periodic in one cell gave similar results.
Only the triplet case was considered but the singlet case
should give similar results \cite{ds94,oppen}.

The results of Fig.1a show that at fixed interaction and strong
disorder $W/V=15$ the $P(s)$ statistics approaches
to the Poisson distribution $(\eta=1)$ at large system size L
and large $r_s= U L/(2 \sqrt{2\pi}V)$.
This means that all states are localized. For smaller disorder
the situation becomes different (Fig. 1b,c). While near
the ground state still $\eta \rightarrow 1$ for large $L$,
the tendency is inverted above some critical energy $\epsilon_c$
where $\eta \rightarrow 0$. All curves $\eta(\epsilon)$
for different $L$ are crossed in one point in a way
similar to the 3D Anderson transition studied in \cite{shklov}.
This result can be understood in the following way.
At strong Coulomb interaction $U \sim V$ 
the excitation energy $\epsilon$
is related to the distance between electrons $R$: 
$\epsilon \sim U/R$ (similar relation was used in \cite{efros}
for the Coulomb glass). At higher $\epsilon$
the distance $R$ becomes smaller, the interaction is stronger
and for $\epsilon > \epsilon_c$ the delocalization
border $R \sim U/\epsilon \sim l_1 r_L^{1/3}$ (\ref{deloc}) 
is crossed and the states 
\vskip -2.0cm
\begin{figure}
\epsfxsize=3.8in
\epsfysize=3.0in
\epsffile{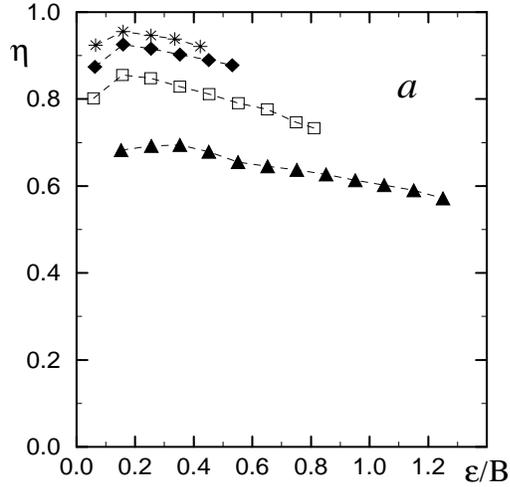}
\vglue 0.2cm
\epsfxsize=3.8in
\epsfysize=3.0in
\epsffile{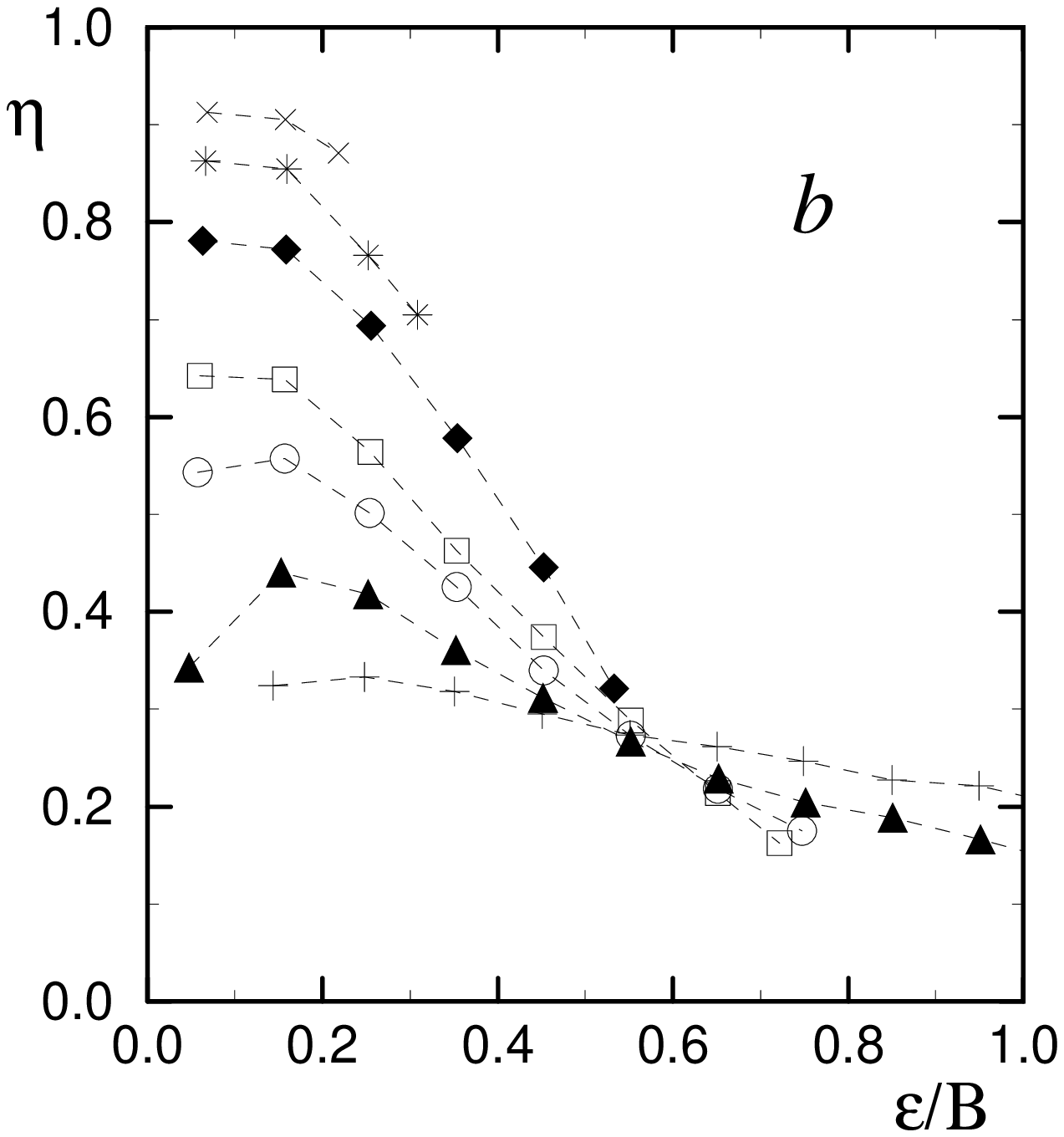}
\vglue 0.02cm
\epsfxsize=3.8in
\epsfysize=3.0in
\epsffile{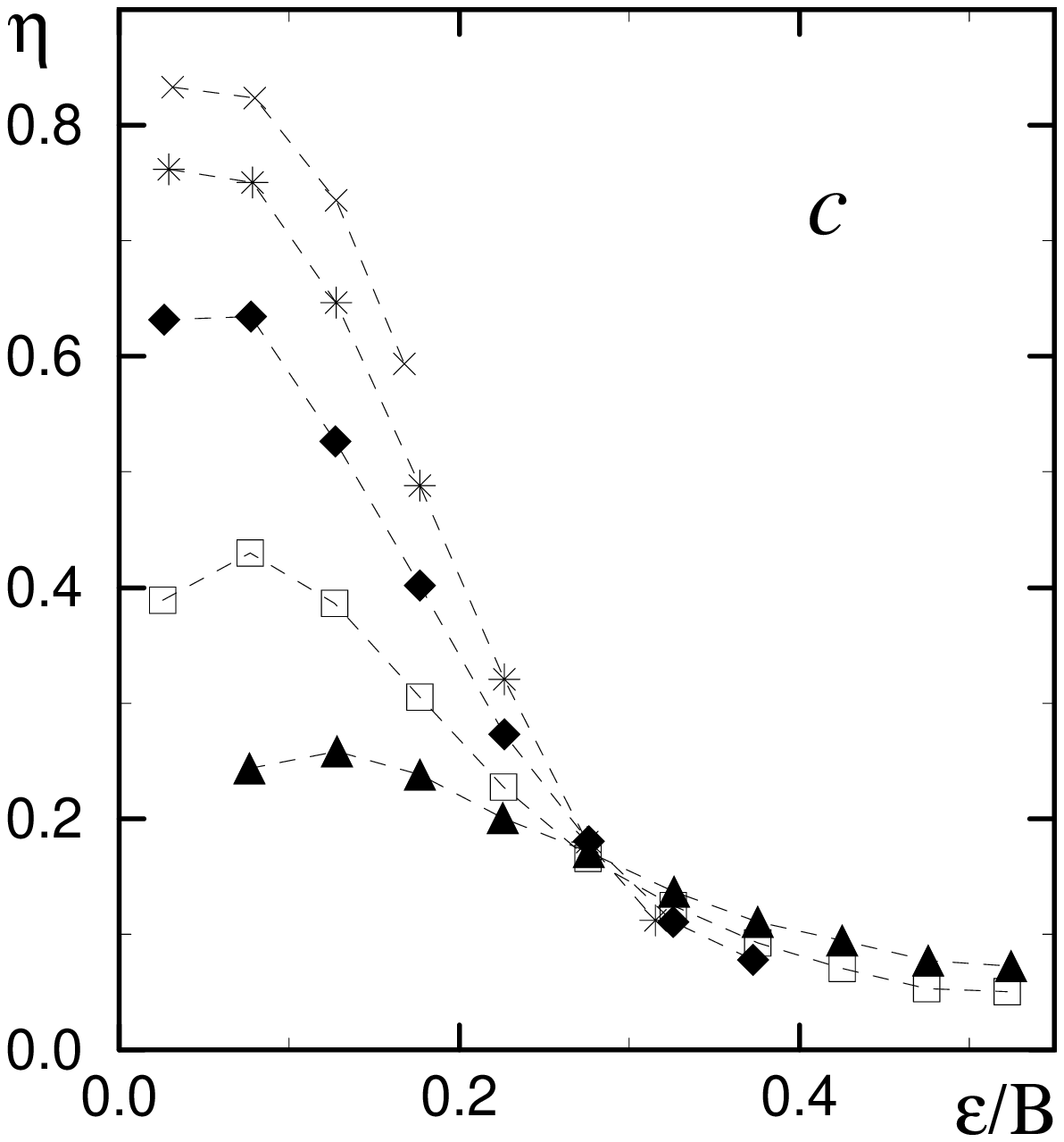}
\vskip -1.0cm
\epsfxsize=3.8in
\epsfysize=3.0in
\epsffile{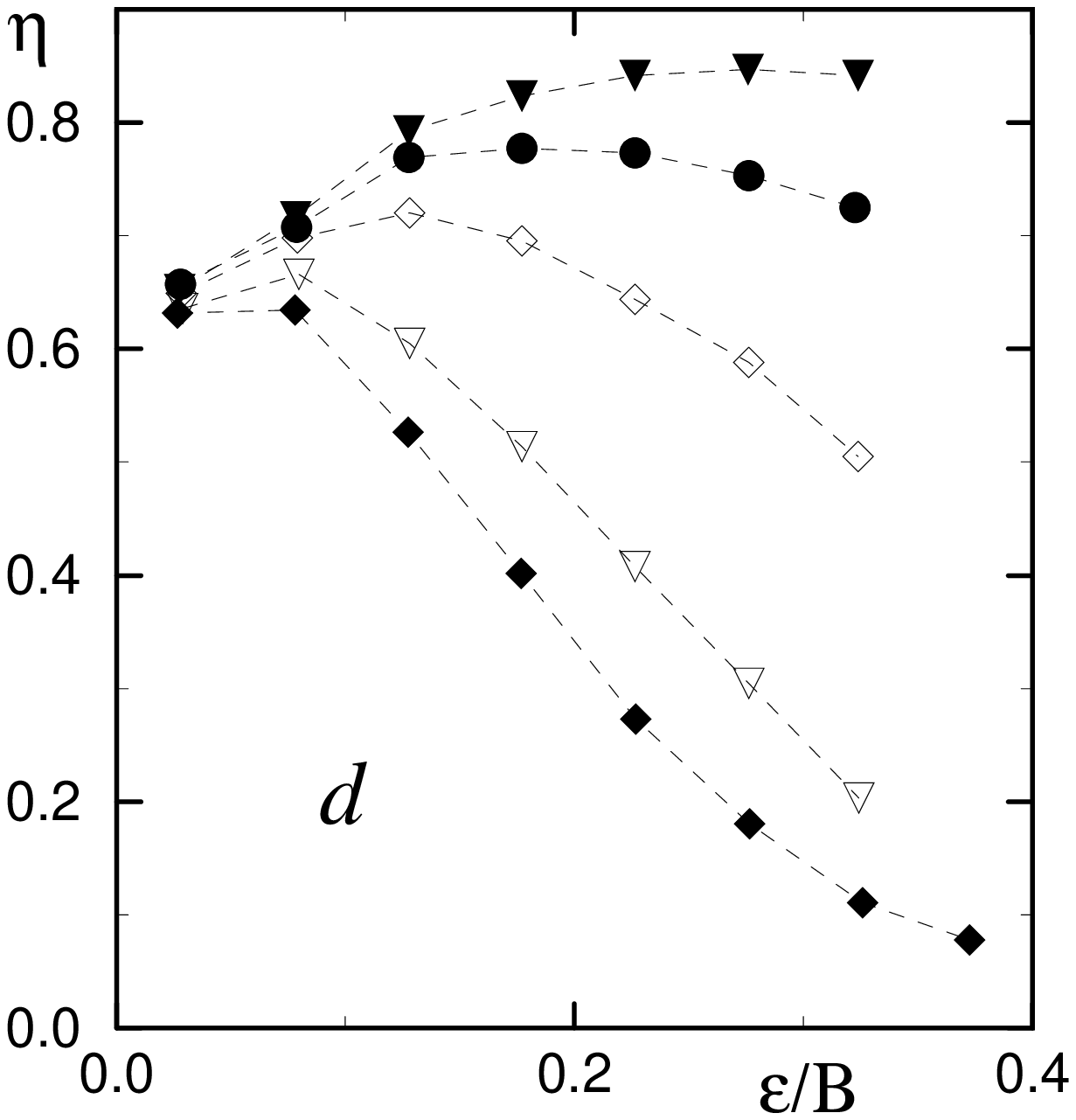}
\vglue 0.02cm
\caption{Dependence of $\eta$ on the rescaled 
one-electron energy $\epsilon/B$ (with $B=4 V$)
for different $W$, system size $L \; (a - c)$ and
interaction strength $U \;(d)$. For $(a - c)$: the size is
$L = 6 (+); 8 $(full triangle); 10 (o);
$12 (\Box)$; 16 (full diamond); 20 (*); 24 $(\times)$,
so that $2.39 \leq r_s \leq 9.57$, $U/V=2$ and
$W/V= 15 (a); 10 (b); 7 (c)$. For $(d): W/V=7, L=16$
and $U/V = 2$ (full diamond); $1 (\bigtriangledown);
0.4 (\diamond)$; 0.2 (full circle); 0.1 (full trangle).}
\label{fig1}
\end{figure}
become delocalized. Since 
the distance $R$ is related
with the two electron energy $E=2 \epsilon \sim U/R$
the spacing statistics $P(s)$, which is local in energy 
and therefore also in $R$,
is not influenced by 
states where particles are far from 
each other. In this sense the situation is different from
the case of short range interaction. According to the above 
arguments $\tilde{\epsilon_c} = \epsilon_c l_1^{4/3}/B$ should
remain constant when $l_1$ changes with disorder. 
The value of $l_1$ can be extracted from the average
inverse participation ratio $\xi_1 = 1/\sum |\psi|^4$ computed 
for one-particle states in the middle of the band
$(l_1 \sim \sqrt{\xi_1})$. For $L=24$ and $W/V=10; 7; 5$ 
we have respectively $\xi_1 = 11.6; 36.7; 84.2$ that with
$\epsilon_c/B \approx 0.6; 0.28; 0.16$ (the case $W/V=5$ is not shown)
gives $\tilde{\epsilon_c} = 3.08\pm0.01$ in a satisfactory
agreement with the above expectations. The variation
of $\eta$ with the interaction $U$ is shown in Fig. 1d.
According to it $\eta$ increases with the decrease of $U$
(states become more localized) in agreement with the general
estimate (\ref{deloc}).
The analysis above allows to understand the dependence
of $\eta$ on $\epsilon$ and $L$. Another reason
for the decrease of $\eta$ at higher $\epsilon$
is related to the fact that the 
two-electron density of states $\rho_2$ grows with
energy that allows to mix levels more easily.
A more detail theory should take this fact into account
but also to analyze the variation of the rate $\Gamma_e$
with $\epsilon$. The results in this direction will be
published elsewhere \cite{moriond2}.

The $P(s)$ statistics  for two electrons in 2D
near the critical point
$\epsilon_c/B$ is shown in Fig. 2. Its comparison
with the critical statistics in 3D Anderson model
taken from \cite{braun} (see also \cite{isa})
demonstrates that both statistics are really 
very close in agreement with the arguments given above.
At the critical point the value of 
$\eta_c$ is close to its value  in the Anderson model
($\eta_c = 0.20$). The small deviations from
this value in the case
of 2D electrons ($\eta_c \approx 0.25 (W/V=10); 0.17 (W/V=7)$)
can be attributed to the fact that the parameter $l_1^{1/3}$ was not
sufficiently large. The investigation of the case 
with larger $l_1$ requires a significant increase 
of the system size $L > 24$.
Indeed, for $L=24$ and $W/V=5$ the localization length
becomes comparable with $L$ ($l_1 \sim \sqrt{\xi_1} \approx 9$)
that gives a decrease of $\eta_c \approx 0.13$. 
\vskip -0.8cm
\begin{figure}
\epsfxsize=3.8in
\epsfysize=3.0in
\epsffile{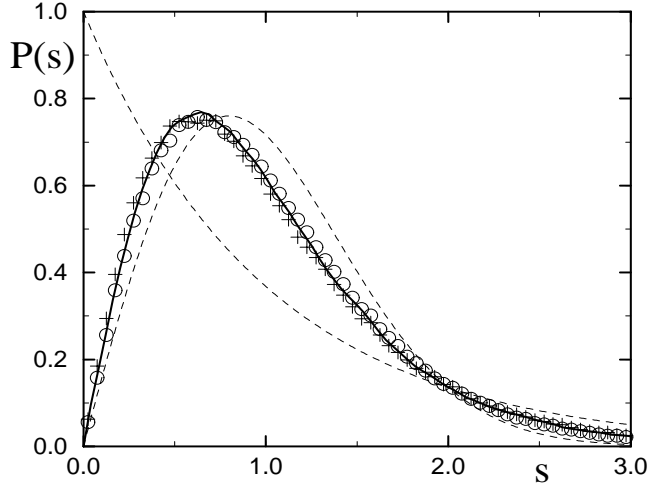}
\vglue 0.2cm
\vskip -0.5cm
\caption{Level statistics $P(s)$ for two 2D electrons
at the critical point: (+) - 
$W/V = 10,  L=12 \; (0.55 \leq \epsilon/B \leq 0.65)$,
total statistics $NS=4 \times 10^5$ (see Fig. 1b);
(o) - $W/V = 7, L=16 \; (0.25 \leq \epsilon/B \leq 0.3)$,
$NS=5 \times 10^5$ (see Fig. 1c). The full line
shows the critical $P(s)$ in 3D Anderson
model ($W/V=16.5, L=14$, taken from Ref. [19]);
the dashed lines give  Poisson statistics
and Wigner surmise.}
\label{fig2}
\end{figure}
Of course, one cannot expect that the simple model
of two electrons considered above will 
explain the variety of experimental results obtained by different groups
\cite{krav96,popovic,canada,yael,alex,pudalov}.
However, it shows some tendencies which are in
agreement with the experiment. Indeed at large
$r_s$ (density lower than some critical $n_c$) 
the experiments demonstrate the
transition from metal to insulator. According
to Fig. 4 in \cite{yael} the density 
at the transition $n_c \propto 1/\sqrt{r_s}$ 
drops exponentially with the increase/decrease of the mobility/disorder 
$\mu \propto 1/W^2$. This qualitatively agrees with the estimate
(\ref{deloc}) according to which near the transition
$\log n_c \sim \log(1/r_s) \sim -\log r_L \sim -1/W^2$.
However, the condition $r_s \gg r_L$ seems to be not well
satisfied and apparently multi-electron effects should be also
taken into account. Another interesting experimental result
(Fig. 2 in \cite{pudalov}) shows that the conductivity $\sigma_c$
near the critical point grows with increase of 
density $n_c$ or disorder $W$.
This is in a qualitative agreement with the estimate (\ref{dif})
according to which $\sigma_c \sim D_e/V \sim 1/l_1^2
\propto r_L^{-2} \propto r_s^{-8/3} \propto n_c^{4/3}$
since near the critical point (\ref{deloc})
$\kappa_e \sim 1$ and $r_s \sim r_L^{4/3}$.
It is also interesting to remark that the 
scaling index $\nu \approx 1.5$ found in \cite{krav96}
is close to the index $\nu \approx 1.5$ near 3D Anderson transition
(the fact that in 3D $\nu \approx s$ can be related to the
observed symmetry of I-V curves).

I thank Y.Hanein and A.Hamilton for the stimulating
discussions of experimental results, D.Braun
for the possibility to use the data of Ref. \cite{braun}
and K.Frahm for a useful suggestion.

\end{document}